\documentclass[twocolumn, prl, amssymb,superscriptaddress, aps,preprintnumbers,amsmath,floatfix]{revtex4}

\usepackage{amsmath}
\usepackage{amssymb}
\usepackage{color,soul}
\usepackage{graphicx}  
\usepackage{mathrsfs}
\usepackage{datetime}
\usepackage[hidelinks]{hyperref}
\usepackage{multirow}
\usepackage{isotope}

\usepackage[dvipsnames]{xcolor}

\usepackage{pdfpages}

\begin{document}

\title{Nuclear Excitation of the  $^{229}$Th Isomer via Defect States in Doped Crystals}

\author{Brenden S. Nickerson}
\email[]{brenden.nickerson@mpi-hd.mpg.de}
\affiliation{Max-Planck-Institut f\"ur Kernphysik, 69117 Heidelberg, Germany}

\author{Martin Pimon}
\affiliation{Center for Computational Material Science, Technische Universit$\ddot{\text{a}}$t Wien, 1040 Vienna, Austria}

\author{Pavlo V. Bilous}
\affiliation{Max-Planck-Institut f\"ur Kernphysik, 69117 Heidelberg, Germany}

\author{Johannes Gugler}
\affiliation{Center for Computational Material Science, Technische Universit$\ddot{\text{a}}$t Wien, 1040 Vienna, Austria}

\author{Kjeld Beeks}

\author{Tomas Sikorsky}

\affiliation{Atominstitut, Technische Universit$\ddot{\text{a}}$t Wien, 1020 Vienna, Austria}

\author{Peter Mohn}
\affiliation{Center for Computational Material Science, Technische Universit$\ddot{\text{a}}$t Wien, 1040 Vienna, Austria}

\author{Thorsten Schumm}
\affiliation{Atominstitut, Technische Universit$\ddot{\text{a}}$t Wien, 1020 Vienna, Austria}

\author{Adriana P\'alffy}
\email[]{Palffy@mpi-hd.mpg.de}
\affiliation{Max-Planck-Institut f\"ur Kernphysik, 69117 Heidelberg, Germany}


\date{\today}

\begin{abstract}
When Th nuclei are doped in CaF$_2$ crystals, a set of electronic defect states appear in the crystal band gap which would otherwise provide complete transparency to  vacuum-ultraviolet radiation. The coupling of these defect states to the 8 eV $^{229m}$Th nuclear isomer in the CaF$_2$ crystal is investigated theoretically.  We show that although previously viewed as a nuisance, the defect states provide a starting point for nuclear excitation via electronic bridge mechanisms involving stimulated emission or absorption using an optical laser. The rates of these processes are at least 2 orders of magnitude larger than direct photoexcitation of the isomeric state  using available light sources. The nuclear isomer population can also undergo quenching when triggered by the reverse mechanism,  leading to a fast and controlled decay via the electronic shell. These findings are relevant for a possible solid-state nuclear clock based on the $^{229m}$Th isomeric transition.
	
\end{abstract}

\maketitle

\newcommand{\ket}[1]{\left| #1 \right\rangle}


Recent years have witnessed the increased interest for a ``nuclear anomaly'' in the actinide region: the first excited state of  the $\isotope[229]{Th}$ isotope lies only 8 eV above the ground state \cite{B.Seiferle2019,  Beck_78eV_2007, Beck_78eV_2007_corrected}. Due to the small transition energy,  this state is long-lived and termed the nuclear isomer $\isotope[229m]{Th}$. The narrow transition line and the possible access with narrow-band vacuum-ultraviolet (VUV) lasers renders $\isotope[229]{Th}$ the first candidate for a nuclear clock of an unprecedented accuracy \cite{Campbell_Clock_2012, Peik_Clock_2003, Peik_Clock_2015, Kazakov_2012}. The $\isotope[229]{Th}$ nuclear clock transition is expected to enable the observation of temporal variations of fundamental constants \cite{Flambaum06, Berengut_ConstVar_PRL_2009, Rellergert_2010} or gravitational shift measurements \cite{Ludlow_OptAtClocks_RevModPhys_2015, Safronova_CommentWense_2016}. Furthermore, it is also a candidate for the first nuclear laser  \cite{Tkalya_2011, Tkalya_Yatsenko_laser_2013}. These applications require a further improvement in the isomer  energy accuracy, at present reported as $E_m = 8.28 \pm 0.17$ eV \cite{B.Seiferle2019}. Furthermore, the development of tunable VUV lasers that would facilitate the search for the $^{229}$Th isomeric transition is at present still  challenging. 

So far, the isomeric transition has been investigated in two experimental approaches: {\bf (i)} using Th ions or atomic beams (see, for instance, Refs.~\cite{Wense_Nature_2016,Thielking2018,B.Seiferle2019}) or {\bf (ii)} in VUV-transparent crystals such as  CaF$_2$ or LiCaAlF$_6$ doped  with $\isotope[229]{Th}$  \cite{Rellergert_2010,Kazakov_2012,Jeet_PRL_2015}. The advantage of the latter approach is that the crystal environment allows for high dopant densities  up to $10^{16}-10^{18}$ cm$^{-3}$, many orders of magnitude larger than those achievable for trapped ions \cite{Kazakov_2012, Stellmer2018, Campbell2011, coulomb_crystal}. This would lead to a larger total  excitation and radiative decay signal of the isomer, which is proportional to the number $N$ of addressed nuclei, and would increase the stability of the potential clock proportionally to $\sqrt{N}$ \cite{PhysRevA.47.3554}. Nevertheless, all attempts of direct isomer excitation with VUV radiation and detection of its radiative decay signal have failed so far \cite{Zhao_2012, PeikZimmermann_Comment_PRL_2013, Jeet_PRL_2015, Zimmermann_thesis_2010, Bilous_UJP_2015}. A reason could be that the explored nuclear transition energy range and the expected transition strength were not accurate, as suggested by more recent results \cite{B.Seiferle2019,Minkov_Palffy_PRL_2017,Minkov_Palffy_PRL_2019}. Furthermore, the reported parasitic background present in Th-doped crystals in the UV and VUV range, such as phosphorescence of crystal defects both intrinsic and laser-induced, or Cherenkov radiation stemming from $\beta$-radioactive daughter nuclei in the $^{229}$Th decay chain \cite{Rellergert_test_2010, Stellmer2018, Stellmer2015, Dessovic_2014, crystaldamage,Zimmermann_thesis_2010} were also a nuisance. In particular, theoretical density functional theory (DFT) predictions  confirm the presence of defect states caused by Th doping within the crystal band gap for  CaF$_2$ in the vicinity of the nuclear transition energy \cite{Dessovic_2014}.

\begin{figure}[htb!]
	\centering
	\includegraphics[scale=1]{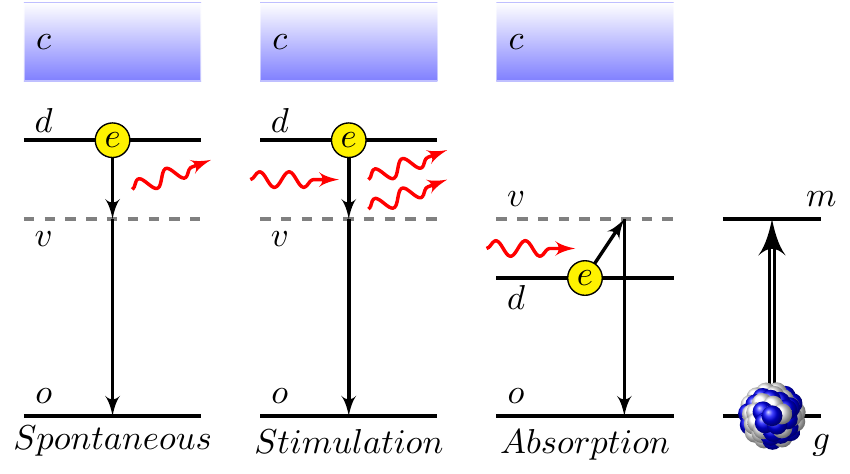}\\
	\caption{EB process for the excitation of $^{229m}$Th from the ground state $|g\rangle$ to the isomeric state $|m\rangle$ (right graph). The initially populated electronic defect states $|d\rangle$ lie in the crystal band gap (between the ground state $|o\rangle$ and the conduction band $|c\rangle$) above or below the isomer energy. The EB process occurs either spontaneously (left graph) or assisted by an optical laser in the stimulated or absorption schemes (middle graphs).  In all cases, EB proceeds via a virtual electronic state $|v\rangle$. See text for further explanations. 
}
	\label{fig:stimEB}
\end{figure}

In this Letter, we consider  a different route to excite the isomeric transition in the VUV-transparent crystal environment exploiting the strong nuclear coupling to the atomic shell. We show that excitation of the defect states in the band gap of CaF$_2$, together with optical laser pumping can lead to an efficient transfer of energy from the crystal electronic shell to the nuclear isomer via a virtual electronic state. This process is a solid-state version of the electronic bridge (EB)  \cite{TkalyaBridge1992, PorsevFlambaum_Brige3+_PRA_2010, PorsevFlambaum_Brige_PRL_2010,Bilous2018,Berengut_PRL_2018,arxiv2020pavlo} studied theoretically so far  only for single ions, typically in ion traps. Our results demonstrate the advantages of this approach: {\bf (i)} it exploits the electronic states of the crystal defects previously viewed as a nuisance, {\bf (ii)} the requirements on the VUV radiation are much reduced and finally {\bf (iii)} the EB nuclear excitation rates are at least 2 orders of magnitude larger than direct photoexcitation of the isomer. Depending on the exact energies of the nuclear transition and of the defect states, we consider both 
an EB mechanism involving stimulated emission as well as a mechanism based on absorption of an optical laser, as illustrated in Fig.~\ref{fig:stimEB}. We also show that the inverse EB process in the crystal can quench the isomeric population producing a 3 orders of magnitude stronger signal than from the mere radiative decay of the nucleus, thus facilitating experimental observation. Controlled quenching could also be relevant for the operation of a solid-state nuclear clock based on the $^{229m}$Th isomeric transition.

For a typical EB process in an ion, the nuclear excitation is accompanied by a decay transition of a different energy in the electronic shell. Due to the energy mismatch, the process involves a virtual electronic level and the emission or absorption of a photon to ensure energy conservation. The EB process in the crystal environment is fundamentally more complicated as it involves the entire crystal electronic shell. We consider in the following the host crystal CaF$_2$ which has an experimentally measured band gap in the region of $11-12$ eV \cite{Rubloff1972, Barth1990, Tsujibayashi2002}. CaF$_2$ displays a cubic lattice structure, in which the dopant thorium in charge state Th$^{4+}$ replaces one of the calcium ions introducing two more interstitial fluorine ions for charge compensation \cite{Dessovic_2014}. DFT calculations using the Vienna Ab initio Simulation Package VASP \cite{PAW, PhysRevB.59.1758} show that there should be eight spin-degenerate defect states $\{|d \rangle \}= \{|d_1\rangle, \ldots, |d_8\rangle\}$ within the $^{229}$Th:CaF$_2$ crystal transparency gap, with energies close in value to the isomer energy. These states are localized on the Th dopant and its $5f$ orbital, while the transition from the crystal ground state $|o\rangle$ to the set $\{|d\rangle\}$ is reminiscent of  a $2p$ orbital electron of an interstitial fluorine ion migrating to the Th ion. We envisage using these $\{|d\rangle\}\rightarrow |o\rangle$ transitions as electronic counterparts to drive the isomer excitation in the EB scheme.  The radiative width of the defect states $\{|d\rangle\}$ modeled by electric dipole ($E1$) decay is $\Gamma^{sp}_{E1}(|d\rangle \rightarrow |o\rangle) \approx 10^6$ s$^{-1}$ and 10 orders of magnitude larger than the isomer width. The defect states are thus much easier to excite by broad-band VUV sources. To this end, we consider an initial electronic state with a uniform population distribution over the defect states $\{|d\rangle\}$, and the final electronic state as the highest energy single spin-degenerate ground state at the Fermi edge corresponding to the ground state $|o\rangle$. 

DFT underestimates the experimentally known band gap of CaF$_2$. In order to match the calculated band gap of the undoped CaF$_2$  crystal with the experimentally measured value of $11.5$ eV, a scaling procedure via the scissors operator is applied in the calculation \cite{Godby88, Levine89}. As a result, the defect states lie in a 0.5 eV interval at approx. $10.5$ eV. The act of scaling introduces uncertainties on the exact energies of the defect states, which thus cannot be undoubtedly assigned above or below the isomer. In the following we will consider both possibilities, as  depicted in Fig.~\ref{fig:stimEB}. The corresponding energy difference $E = \hbar|\omega_{mg}-\omega_{do}|$ between the nuclear isomeric transition ($\hbar\omega_{mg}$)  and  the electronic decay transition of one defect state ($\hbar\omega_{do}$) lies for both cases in the optical domain, where tunable lasers with spectral intensities on the order of $I_{opt} \approx 1$ W/(m$^2$s$^{-1}$) or more are available. If $\omega_{do}>\omega_{mg}$, a laser photon with energy $E$ can assist via a stimulated emission transition to the virtual state  leading to subsequent EB excitation of the isomeric state. In the opposite case where $\omega_{do}<\omega_{mg}$, absorption of an optical laser photon with energy $E$ would render possible the EB process. We consider the laser tuned on the respective resonance and neglect detuning.

The rate  $\Gamma^{st}(a\rightarrow b)$ of a stimulated generic process $|a\rangle\rightarrow |b\rangle$ can be related to the rate of the corresponding spontaneous  process $\Gamma^{sp}(a \rightarrow b)$
 in SI units as \cite{LL_QED_1982, Sobelman_book_1979}
	\begin{align}
		\Gamma^{st}(a\rightarrow b) &= \Gamma^{sp}(a \rightarrow b) \frac{\pi^2c^2\hbar^2}{E^3} I\, ,
		\label{eqn:first}
	\end{align}
where the spectral intensity of the laser source is $I$ in W/(m$^2$s$^{-1}$), with the required photon energy $E=\hbar\omega_{ab}$, and $c$ stands for the speed of light. Via detailed balance, the stimulated rate $\Gamma^{st}(a\rightarrow b)$ can be related to the inverse absorption process rate via $ \Gamma^{ab}(b\rightarrow a) = \Gamma^{st}(a\rightarrow b) \delta(a\rightarrow b)$, with  $\delta(a\rightarrow b) = N_a/N_b$ the ratio of multiplicities $\{|a\rangle\}$ versus $\{|b\rangle\}$.

We now address the considered case of spontaneous EB in the solid-state environment as illustrated in Fig.~\ref{fig:stimEB}, adapting the standard theoretical approach for single ions \cite{TkalyaBridge1992, PorsevFlambaum_Brige3+_PRA_2010, PorsevFlambaum_Brige_PRL_2010,Bilous2018}. The EB process is described by the amplitude of an electric dipole transition for the emission of a photon, occurring together with an electronic transition, both via a set of intermediate states around the virtual state. These processes are simultaneously  accompanied by  nuclear excitation in the transition $|g\rangle\rightarrow |m\rangle$. Photon emission and electronic transition can occur in either order, with only one of them being depicted in Fig.~\ref{fig:stimEB}. The spontaneous EB rate $\Gamma^{sp}_{EB}(|g,d\rangle\rightarrow |m,o\rangle)$ starting from  the set of defect states $\{|d\rangle\}$ can be written in atomic units as
	\begin{align}
		\Gamma^{sp}_{EB}&(|g,d\rangle\rightarrow |m,o\rangle) = \nonumber \\
		&\frac{4}{3} \frac{1}{N_g N_d}\sum_{\substack{m,g,\\d,o}}\left(\frac{\omega_{do}-\omega_{mg}}{c}\right)^3|\langle m,o|\widetilde{\boldsymbol D}_{E1}|g,d\rangle|^2\, ,
	 \label{eqn:EBsponexc1_cry}
	\end{align}
where the states  $|m,o\rangle = |m\rangle|o\rangle$ are characterized by the quantum numbers defining the nuclear state, in this case the isomer $m$  and the quantum numbers of the electronic state, namely the  electronic ground state $o$.  The number of initial nuclear and participating electronic states in the sets $\{|g\rangle\}$ and $\{|d\rangle\}$ are $N_g$ and $N_d$ respectively. In the crystal environment,   we  average over the transitions between sets of nondegenerate electronic levels in the ground state and defect bands. This is accounted for by the summations over the sets $\{|d\rangle\}$ and  $\{|o\rangle\}$  in Eq.~\eqref{eqn:EBsponexc1_cry}.

The effective $E1$ bridge operator matrix element for the crystal environment reads
	\begin{widetext}
		\begin{align}
		\langle m,o|\boldsymbol{\widetilde{D}}_{E1}|g, d\rangle &= \sum_{\lambda K,q}(-1)^q\left[\sum_{n} \frac{\langle o|\boldsymbol{D}_{E1}|n\rangle\langle n|\mathcal T_{\lambda K,q}|d\rangle}{\omega_{dn}-\omega_{mg}}
		+\sum_{k} \frac{\langle o|\mathcal T_{\lambda K,q}|k\rangle\langle k|\boldsymbol{D}_{E1}|d\rangle}{\omega_{ok}+\omega_{mg}}\right]\langle m|\mathcal M_{\lambda K,-q}|g\rangle\, . \label{eqn:EBsponexc2}
		\end{align}
	\end{widetext}
Here, $\lambda K$ represent the multipolarities of the electronic and nuclear spherical tensors $\mathcal{T}_{\lambda K,q}$ and $\mathcal M_{\lambda K,-q}$ and $q= (-K, -K+1,\ldots, K-1,K)$ are the spherical components thereof  \cite{Akhiezer_QED,Varshalovich_QTAM}. Due to the presence of the virtual state, the matrix element contains summations over intermediate electronic states denoted by $|n\rangle$ and $|k\rangle$ in Eq. \eqref{eqn:EBsponexc2} which include all neighbouring unoccupied electronic states.

The electric dipole operator in Eq.~\eqref{eqn:EBsponexc2} describes the emission or absorption of a photon and  is given by $\boldsymbol{D}_{E1} = -\boldsymbol r$, where $\boldsymbol r$ is the position relative to the thorium nucleus which is considered the origin. The isomeric transition in $^{229}$Th is a magnetic dipole ($M1$) transition with an electric quadrupole ($E2$) admixture \cite{PavloE2}. This restricts the sum over $\lambda K$ to these two multipolarities. 
	In the nonrelativistic limit, the magnetic-dipole coupling operator reads \cite{Abragam}
	\begin{align}
	\mathcal{T}_{M1,q}&= \frac{1}{c}\left[\frac{l_q}{r^3}-\frac{\sigma_q}{2r^3} + 3\frac{r_q (\boldsymbol \sigma \cdot \boldsymbol r)}{2r^5} + \frac{4\pi}{3}\sigma_q \delta(\boldsymbol r)\right],
	\label{eqn:TM1}
	\end{align}
where $\boldsymbol l$ ($l_q$) is the orbital angular momentum of the electron and $\boldsymbol \sigma$ ($\sigma_q$) are the Pauli matrices (in spherical basis). The electric-quadrupole coupling operator is given by \cite{Varshalovich_QTAM}
	\begin{align}
	\mathcal{T}_{E2,q} &= -\frac{1}{r^3} \sqrt{\frac{4\pi}{5}}Y_{2,q}(\theta, \phi),
	\end{align}
	where $Y_{2,q}$ are the spherical harmonics. Last, $\langle m|\mathcal M_{\lambda K,-q}|g\rangle$ are the matrix elements of the nuclear transition operators which are obtained via the Wigner-Eckart theorem \cite{Edmonds_AM} using the theoretical reduced transition probabilities   $B_\downarrow$ in  Weisskopf units (W.u.) $B_W(M1, m\rightarrow g) = 0.0076 \,\,\,\text{W.u.}$, $B_W(E2, m\rightarrow g) = 27.04 \,\,\,\text{W.u.}$  \cite{Minkov_Palffy_PRL_2017}.


For the electronic wave functions, ab-initio calculations were carried out in VASP \cite{PAW, PhysRevB.59.1758} using Perdew-Burke-Ernzerhof  \cite{PBE} and Heyd-Scuseria-Ernzerhof  \cite{HSE} functionals, where we reconstructed the all-electron Kohn-Sham wave functions from the projector augmented wave method \cite{Blochl1994}. Eq.~\eqref{eqn:EBsponexc2} requires the knowledge of energies and wave functions for the ground state, defect and conduction bands. Some of these electronic states are localized tightly around thorium and others have their origin elsewhere in the crystal's unit cell or are delocalized. We have used a unit cell of 66 fluorine, 31 calcium, and a single thorium atom for the VASP calculations. Compared to atomic EB calculations,  the  wave functions of electrons in the crystal environment are not necessarily eigenstates of either angular momentum or parity. The spatial parts of the wave functions are therefore   only defined by their energy.	This raises additional numerical challenges as selection rules cannot be used for analytical simplifications of the computation. Wave functions were calculated on a spherical grid where the number of points is $(N_r, N_\theta, N_\phi) = (353,29,60)$, the spacing in angular components is constant and the spacing in the radial component follows $r_n=r_0e^{n/\kappa}$ with $r_0=0.000135\,a_0$, $a_0$ is the Bohr radius, and $\kappa = 31.25$. Spherical grids as large as $(N_r, N_\theta, N_\phi) = (353,44,90)$  were tested but did not improve the accuracy of the result significantly. 

Our numerical results for the  spontaneous EB rate considering the DFT defect states energies around 10.5 eV yield  $\Gamma^{sp}_{EB}(|g,d\rangle\rightarrow |m,o\rangle) \approx 2.5\times 10^{-8}$ s$^{-1}$. We have tested the rate convergence  by increasing the number of included intermediate spin-degenerate conduction band states up to 230. The conduction band $\{|c\rangle\}$ offers an infinite intermediate-state set  for the $n$ and $k$ summations in Eq.~\eqref{eqn:EBsponexc2}, and the corresponding denominators are only slowly suppressing their contributions. Our test shows that convergence is achieved and the order of magnitude of the rate is stable throughout the entire tested range.

The  EB nuclear excitation schemes  require initial population of the defect, i.e., prior excitation of an electron from the ground state  to one of the defect states $|o\rangle \rightarrow \{|d\rangle\}$. This initial excitation can be achieved with VUV sources, for instance, available VUV lamps \cite{Stellmer2018} with $N \approx 3$ photons/(s$\cdot$Hz), a focus of $f=0.5$ mm$^2$ which gives $I ={N\hbar\omega_{do}}/(2\pi f) \approx 1.6 \times 10^{-12}$ W/(m$^2$ s$^{-1}$) and a FWHM linewidth of $\approx 0.5$ eV, or alternatively with the Advanced Light Source (ALS) in Berkeley, USA, described in Ref.~\cite{Rellergert_2010} with spectral intensity $I \approx 0.5 \times 10^{-8}$ W/(m$^2$ s$^{-1}$). The steady-state total defect population $\rho_d$ can be calculated considering the stimulated ($st$) and spontaneous ($sp$) $E1$ decay rates and the VUV-photon absorption ($ab$) processes for the process $|o\rangle \leftrightarrow \{|d\rangle\}$
 	\begin{align}
		\dot{\rho}_d = \rho_o \Gamma^{ab}_{E1} - \rho_d \left(\Gamma^{sp}_{E1} + \Gamma^{st}_{E1}\right)\, ,
	\label{eqn:popc}
	\end{align}
where $\rho_o$ is the occupation probability of the ground state,  $\rho_o = 1-\rho_d\simeq 1$. The steady state solution $\dot{\rho}_d=0$ gives the approximate occupation probability of the defect, which yields $\rho_d \approx 2.7\times 10^{-8}$ for the VUV lamp and $\rho_d \approx 10^{-4}$ for ALS  parameters.

\begin{figure}[h!]
		\centering
		\includegraphics[scale=1,page=1]{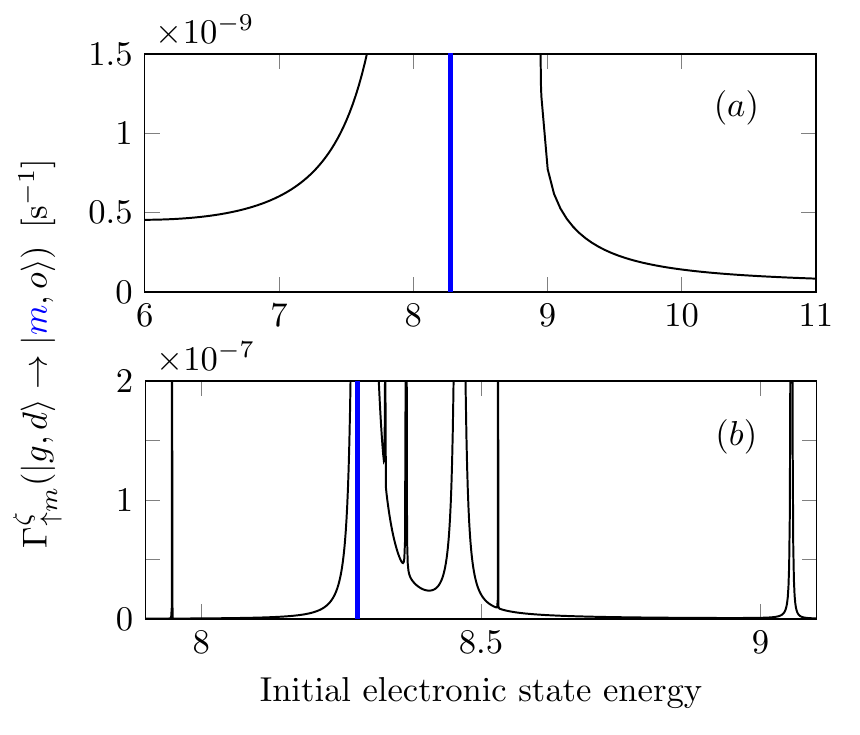}
		\caption{EB excitation rates via absorption ($\zeta=ab$) and stimulated ($\zeta=st$) process as a function of the initial electronic state energy. $(a)$ Average driven EB process where the initial state is the set of all defect states $\{|d\rangle\}$. $(b)$ Driven EB process for the initial state $|d\rangle = |d_7\rangle$. The isomeric energy is shown as a vertical blue line at $8.28$ eV. }
		\label{fig:CBst}
	\end{figure}

We now turn to the total rate of isomer excitation starting from the electronic shell in one of the defect states via both  spontaneous and  laser-induced channels. When the defect states have higher energy than the isomer $\omega_{do}>\omega_{mg}$, the total rate of the isomer excitation for both stimulated and spontaneous EB processes can be written as
	\begin{align}
		\Gamma_{\uparrow m}^{st}=\rho_g\rho_d &\left[\Gamma^{sp}_{EB}(|g,d\rangle\rightarrow |m,o\rangle) \right. \nonumber\\
		&\,\,\,\,\, \left. +\,\, \Gamma^{st}_{EB}(|g,d\rangle\rightarrow |m,o\rangle)\right]\, , 
	\label{eqn:stimtot}
	\end{align}
where $\rho_g$ is the  occupation probability of the nuclear ground state, which we consider initially to be unity. In the opposite case $\omega_{do}<\omega_{mg}$, the EB process requires laser absorption and the total nuclear excitation rate can be written as
	\begin{align}
		\Gamma_{\uparrow m}^{ab} = & \,\rho_g\rho_d\Gamma^{sp}_{EB}(|g,d\rangle\rightarrow |m,o\rangle) \frac{\pi^2c^2\hbar^2}{E^3} I_{opt} \\ \nonumber 
		&\times 
	 	\delta(|g,d\rangle\rightarrow |m,o\rangle)\, . \label{eqn:absortot}
	\end{align}
Using the last two expressions we obtain the rates for both stimulated and absorption EB processes as a function of the defect energy illustrated in Fig.~\ref{fig:CBst}. The energy of the defect states in $\{|d\rangle\}$ is varied as a group by subtracting the same constant from each state energy. For the calculation we have used $I_{opt} =1$ W/(m$^2$s$^{-1}$) for the optical laser and $I_{VUV} = 1.6 \times10^{-12} $ W/(m$^2$s$^{-1}$) for the VUV source that originally excites the defect states. We obtain EB rates $\Gamma_{\uparrow m} \gtrsim 10^{-10}$ s$^{-1}$ for both the stimulated and absorption schemes.
 
Figure \ref{fig:CBst}(a) shows the rate $\Gamma_{\uparrow m}^{\zeta}$ for the driven EB process averaged over all eight spin-degenerate defect states $|d\rangle = \{|d\rangle\}$,  as a function of the average energy of the defect states in the range 6--11 eV. Both the stimulation ($\zeta=st$)  and the absorption ($\zeta=ab$)  schemes are considered. A clear resonance of the form $\Gamma^{\zeta}_{\uparrow m}\propto 1/(\omega_{mg}-\omega_{do})^2$ is visible as expected from the denominators of Eqs.~\eqref{eqn:EBsponexc1_cry} and \eqref{eqn:EBsponexc2}.   Both EB schemes are upwards of two orders of magnitude faster than directly driving the isomeric transition with the VUV source alone. This is a significant improvement for the isomer excitation in Th:CaF$_2$. When approaching the isomer energy from a distance the average rates of Fig.~\ref{fig:CBst}(a) provide a general view of the resonant structure. In the region of the isomer energy one must consider specific initial states individually. To gain a better understanding of this region, we plot in Fig.~\ref{fig:CBst}(b) the resonant structure of the EB rate corresponding to the initial defect state $|d_7\rangle$, which has the largest radiative width and can be most efficiently populated. Here we see a variety of resonances corresponding to cases when the intermediate virtual state comes close to one of the other defect states. Near these resonances the rate of excitation can be orders of magnitude larger than discussed so far. The exact comb structure will depend on the chosen initial state as well as the precise energies of the other surrounding defect states, which are yet to be determined in  future experimental measurements.  The latter  can be performed in $^{232}$Th:CaF$_2$ crystals with the advantage of less radioactive background. 

The laser-assisted bridge mechanisms can also be used to quench an already generated isomeric state population.  Quenching refers here to the conversion of nuclear isomeric population into electronic defect population. The subsequent defect decay is 10 orders of magnitude faster than the radiative decay of the nuclear isomer. Considering the case $\omega_{do}>\omega_{mg}$ with defect states energies around 10.5 eV, the optical laser induces a 0.07 s$^{-1}$ deexcitation of the isomer via the stimulated EB process, which is nearly 3 orders of magnitude faster than the spontaneous radiative decay of the isomer estimated to be $\Gamma_\gamma \approx10^{-4}$ s$^{-1}$ \cite{Minkov_Palffy_PRL_2017}. With occupation probabilities calculated for initial excitation of the isomeric state via the above outlined stimulated EB method, the quenching process has a rate of $\Gamma_{\downarrow m} \approx 10^{-10}\,\,{\text s}^{-1}$. The total number of decay photons one would measure is of the order $N \Gamma_{\downarrow m}$ where $N$ is the total number of nuclei in the crystal exposed to the initial stimulated EB excitation scheme. For a $1$ cm$^3$ crystal one could have  $N\gtrsim 10^{14}$ thorium nuclei giving a signal of $\approx 10^4$ decay photons per second, with the emitted photon energy coinciding with the electronic defect energies. This signal is 3 orders of magnitude larger than the radiative decay of the nucleus considering the same excitation parameters. Quenching could therefore be advantageous for a faster clock interrogation scheme in the crystal environment.

	
In conclusion, we have presented a new method for isomer excitation in the crystal environment via the coupling to the electronic states using a defect  in $^{229}$Th:CaF$_2$. Our calculations show that excitation rates via both stimulated and absorption EB processes starting from  previously excited defect states are at least 2 orders of magnitude larger than direct photoexcitation using the same VUV source. In addition, the isomer coupling to the defect states could be used for an efficient quenching of the nuclear transition and controlled isomer decay. Our findings support experimental efforts toward a solid-state nuclear clock employing Th-doped VUV-transparent crystals.
	
	
We gratefully acknowledge funding from the European Union's Horizon 2020 research and innovation programme under Grant agreement No. 664732 ``nuClock''. This work is part of the ThoriumNuclearClock project that has received funding from the European Research Council (ERC) under the European Union’s Horizon 2020 research and innovation programme (Grant agreement No. 856415). The project has also received funding from the EMPIR programme cofinanced by the Participating States and from the European Union's Horizon 2020 research and innovation programme.	
	
\bibliographystyle{unsrt}
\bibliography{references}

\end{document}